\pdfoutput=1
\documentclass[11pt]{arxiv-article}
\usepackage{standalone}
\RequirePackage{tikz}
\usetikzlibrary{calc}

\bibliography{References}

\usepackage{Definitions}

\usepackage{pifont}
\usepackage{hyperref}
\usepackage{mathrsfs}

\allowdisplaybreaks

\newcommand{\OfficialTitle}{
  Following the flow for large N and large charge
}
\title{\setstretch{1.4}
  {\color{Thoughtless}\Huge\textbf{\dosserif\OfficialTitle}}
}

\hypersetup{pdfauthor={Domenico Orlando, Susanne Reffert, Tim Schmidt},pdftitle={\OfficialTitle}}

\author{%
  \begin{minipage}{.97\linewidth}
    \vspace{1cm}
    \begin{center} \dosserif%
      {\small
         \textbf{Domenico Orlando}\textsuperscript{\ding{72}\ding{73}},
         \textbf{Susanne Reffert}\textsuperscript{\ding{73}} and
         \textbf{Tim Schmidt}\textsuperscript{\ding{73}} 
         }
    \end{center}
    \vspace{1cm}
      \authorBlock{\ding{72}}{\dosserif{} INFN sezione di Torino,
       via Pietro Giuria 1, 10125 Turin, Italy}
     \authorBlock{\ding{73}}{\dosserif{} Albert Einstein Center for Fundamental Physics\\
       Institute for Theoretical Physics, University of Bern,\\
       Sidlerstrasse 5, CH-3012 Bern, Switzerland}
  \end{minipage}
}

\date{}

\begin{document}

\setstretch{1.2}

\numberwithin{equation}{section}

\begin{titlepage}

  \maketitle

  \thispagestyle{empty}

  \vfill\dosserif{}

  \abstract{\normalfont{}\noindent{}%
    We discuss the $O(2N)$ vector model in three dimensions.
    While this model flows to the Wilson--Fisher fixed point when fine tuned, working in a double-scaling limit of large N and large charge allows us to study the model away from the critical point and even to follow the RG flow from the UV to the IR.
    The crucial observation is that the effective potential --- at leading order in N but exact to all orders in perturbation theory --- is the Legendre transform of the grand potential at fixed charge.
    This allows us to write an effective action and the free energy for generic values of the coupling in a very simple fashion and without evaluating any Feynman diagrams.
  }

\vfill

\end{titlepage}

\setstretch{1.2}

\tableofcontents

\section{Introduction}
\label{sec:intro}

Working in a sector of fixed and large global charge leads to numerous simplifications in strongly coupled models (for a review, see~\cite{Gaume:2020bmp} and references therein). While most of the large-charge literature focuses on \acp{cft}, the approach is in principle applicable to any  quantum field theory. 

The O(2n) vector model in three dimensions has been studied in great detail at large charge~\cite{Hellerman:2015nra,Alvarez-Gaume:2016vff,Monin:2016jmo,Loukas:2017lof,Badel:2019oxl,Badel:2019khk,Alvarez-Gaume:2019biu,Dondi:2021buw,Antipin:2020abu,Antipin:2021jiw,Jack:2021ypd}. While these works all study the conformal \ac{ir} \ac{wf} fixed point, we use here the combined power of working at large charge and large N to study the model away from the fixed point for generic values of the quadratic and quartic couplings. 

While working at large N gives a lot of control, working in the double-scaling limit of large charge at fixed \(Q/N\) provides us with essential extra structure.
From the grand potential (which is a function of the chemical potential dual to the fixed charge), computed at leading order in N, we can directly read off an effective \ac{nlsm} action.
This large-$Q$ effective action captures the system at large charge, it describes the spontaneous symmetry breaking due to the fixed charge and contains the same type of terms as the \ac{nlsm} at finite N, see \emph{e.g.} the result for the O(2) model~\cite{Hellerman:2015nra}.
The large-N behavior of the model can be described through an effective action which takes the form of a \ac{lsm}.
It can be studied both in the limit of large and small charge, and captures the flow from the \ac{uv} to the \ac{ir}.
The two actions apply in general to different ranges of the parameters $N$ and $Q$ but must agree in the double scaling limit of $Q$, $N$ large and $Q/N$ fixed.

Using this agreement we find that the grand potential is related to the effective potential by a Legendre transform in the  \ac{vev} of the collective field, $m^2$.
This observation is crucial.
The resulting effective potential is exact in the sense that it corresponds to a resummed expression which includes infinitely many  Feynman diagrams, each computing the leading-\(N\) contribution to a monomial in the expansion of the effective potential.
This is due to the large-N limit which captures directly the quantum effects in terms of a semiclassical action.
In the massless case we indeed reproduce a known result found diagrammatically in~\cite{Appelquist:1982vd}.

Applied to the case of the $\varphi^4$ model on flat space the grand potential at leading order in N is given by two terms only and leads to a closed expression for the effective potential.
Depending on whether the coefficient $r$ of the quadratic term in the tree-level potential is bigger, equal, or smaller than zero we are physically in the unbroken phase, at criticality or in the broken phase.
The power of the Legendre transform shows itself in the case of the broken phase where the double-well tree-level potential is replaced by a convex effective potential with a flat line between the original minima.

All these results are obtained in a very simple and transparent fashion and without the need of evaluating and resumming  Feynman diagrams.
Also the free energy in the different phases can be computed straightforwardly via another Legendre transform, this time in $m$.

\medskip

Using the same methodology that is introduced here, the O(2N) vector model is studied in various dimensions in~\cite{other-paper}, where the criterion of convexity allows us to distinguish between unitary and non-unitary theories. 

\bigskip
The plan of this paper is as follows.
In Section~\ref{sec:OneLoopFixedCharge}, we extend the large N results of~\cite{Alvarez-Gaume:2019biu} to non-critial values of the quadratic coupling \(r\) and of the quartic coupling $u$ and compute both the expressions for the free energy and for the grand potential at large charge.
In Section~\ref{sec:EFT} we observe that the effective potential at leading order in N is given by the Legendre transform of the grand potential.
It is important to use the definition of the Legendre transform that includes taking the supremum.
After the general remarks of Section~\refstring{sec:largeN}, we explicitly study the $\varphi^4$ model on flat space in Section~\refstring{sec:phi4}.
Depending on the value of the coefficient $r$ of the quadratic term, we distinguish three cases in Section~\ref{sec:torusEFT}: the unbroken phase ($r>0$), the critical point ($r=0$) and the broken phase ($r<0$).
For each of them, we compute the effective potential and observe that unlike the tree-level potential, it is always convex.
In Section~\ref{sec:freeEnergyFlat}, we give the free energy  density for the three cases.
In Section~\ref{sec:NonperturbativeFlat} we derive the first nonperturbative correction to the grand potential and its Legendre transform, the effective potential.
In Section~\refstring{sec:sphereEFT} we treat the critical theory on the sphere.
Since we only have a series expansion for the zeta function, we do not get a closed expression for the effective potential, but give the first few terms of $V$ as an expansion in the curvature.
In Section~\refstring{sec:Conclusions}, we give conclusions and an outlook.

\section{The effective action from large N}
\label{sec:largeN}

We want to discuss the Landau--Ginzburg model for $2N$ real scalar fields in the vector representation of $O(2N)$ in three dimensions.
When fine-tuned, this model flows to the \ac{wf} fixed point in the \ac{ir}.
In the following we will rely on the same ideas as~\cite{Alvarez-Gaume:2019biu} and compute the one-loop results at fixed charge in the double scaling limit \(N \to \infty\), \(Q \to \infty\) with \(Q/N\) fixed, while keeping the coupling to the $\abs{\phi}^4$ term finite. 

\subsection{One-loop result at fixed charge}
\label{sec:OneLoopFixedCharge}

We work in Euclidean signature on $S^1_\beta \times \Sigma$, where $\Sigma$ is a Riemann surface. In the following we will specialize to the cases of flat space (the torus $T^2$) and the sphere $S^2$. We start with the action for $N$ complex scalar fields $\phi_i$,
\begin{align}\label{eq:SThetaphi4}
    S[\phi_i] = - \int \dd{\tau}\dd{\Sigma} \left[g^{\mu \nu}(\del^i_\mu \phi_i)^\dag (\del^i_\nu \phi_i) - r(\phi_i^\dag\phi_i) - \frac{u}{2N}(\phi_i^\dag\phi_i)^2\right].
\end{align}
For $r=R/8$, where $R$ is the scalar curvature of $\Sigma$, the model flows to a non-trivial conformal fixed point in the \ac{ir} where $u$ diverges.
Here we want to explore the regime where $u$ is finite.

After a Stratonovich transformation we have an action in terms of $\phi_i$ and the field $\lambda$~\cite{stratonovich1957method,Hubbard:1959,Moshe:2003xn},
\begin{align}\label{eq:STheta}
    S[\phi_i,\lambda] = - \int \dd{\tau}\dd{\Sigma} \left[g^{\mu \nu}(\del^i_\mu \phi_i)^\dag (\del^i_\nu \phi_i) - (r+\lambda)(\phi_i^\dag\phi_i) - \frac{\lambda^2}{2u}\right].
\end{align}

We want to study this system at fixed charge, where the canonical partition function is given by
\begin{equation}%
  \label{eq:canonicalPartitionFn}
  \begin{aligned}
  Z(Q_1, \dots, Q_N) &= \Tr[ e^{-\beta H} \prod_{i=1}^N \delta(\hat Q_i - Q_i)]
   = \int_{-\pi}^{\pi} \prod_{i=1}^N \frac{\dd{\theta_i}}{2 \pi} \prod_{i=1}^N e^{i \theta_i Q_i} \Tr[ e^{-\beta H} \prod_{i=1}^N e^{- i \theta_i \hat Q_i}] \\
  & = \int_{-\pi}^{\pi} \prod_{i=1}^N \frac{\dd{\theta_i}}{2 \pi} \prod_{i=1}^N e^{i \theta_i Q_i} Z_{gc}(i\theta_1, \dots, i\theta_N),
\end{aligned}
\end{equation}
where $\hat Q_i$ are the Noether charges corresponding to the Cartan generators of the global $O(2N)$ symmetry and \(\theta_i\) are the dual imaginary chemical potentials in the grand canonical description.

It is convenient to introduce the covariant derivative $D_\mu$, defined as
\begin{align}
    D^i_\mu \phi_i := \begin{cases}
		\left(\partial_0 + i \frac{\theta_i}{\beta}\right)\phi_i, & \mu = 0, \\
		\partial_i \phi_i, & \text{otherwise,}
	\end{cases}
\end{align} 
so that the chemical potentials are realized by substituting \(\del_\mu \) with \(D_\mu\) in the action.
As shown in~\cite{Alvarez-Gaume:2016vff}, the path integral localizes around a trajectory that only depends on the sum of the charges \(Q = Q_1 + \dots + Q_N\).

We will decompose the fields $\phi_i$ into a constant part $A_i/\sqrt{2}$ and a fluctuating part $u_i$. The same applies to the field $\lambda$ where we choose the constant part to be $m^2-r$ and the fluctuating part $\hat{\lambda}$. Up to a total derivative the action becomes
\begin{equation}
\begin{aligned}
    S_Q[u_i,\hat{\lambda}] ={}& \sum_{i=1}^N \biggl[-i\theta_i Q_i + 	\int \dd{\tau}\dd{\Sigma} \biggl[(D^i_\mu u_i)^\dag (D_i^\mu u_i)	+ \frac{\theta_i^2 A_i^2}{2\beta^2} \\ 
    &+ (m^2+\hat{\lambda})\left(\frac{A_i^2}{2} + \frac{A_i}{\sqrt{2}}(u_i^\dag + u_i) + u_i^\dag u_i\right)	\biggr] -\left(\frac{(m^2-r)^2}{2u} + \frac{\hat{\lambda^2}}{2u} + \frac{(m^2-r)\hat{\lambda}}{u}\right)\biggr].
\end{aligned} 
\end{equation}
Since the $u_i$ appear at most quadratically we can integrate them out and get
\begin{align}
    Z = \int \prod_{i=1}^N\frac{d\theta_i}{2\pi} \DD{u_i} \DD{\hat{\lambda}} e^{-S_Q[u_i,\hat{\lambda}]} = \int \prod_{i=1}^N\frac{d\theta_i}{2\pi} \mathcal{D}\hat{\lambda} e^{i\theta_i Q_i}e^{-S_\theta^{\textrm{eff}}[\hat{\lambda}]},
\end{align}
where 
\begin{equation}
	\begin{aligned}
	    S_\theta^{\textrm{eff}}[\hat{\lambda}] ={}& \sum_{i=1}^N \biggl[	V_{\Sigma}\beta \left(\frac{\theta_i^2}{\beta^2}+ m^2 - \frac{(m^2-r)^2}{u A_i^2}	\right)\frac{A_i^2}{2} + \Tr\log(-D_\mu D^\mu + m^2 + \hat{\lambda}) \\ 
	&- \frac{A_i^2}{2}\Tr(\hat{\lambda}\Delta^i\hat{\lambda})  
	- \frac{1}{u}\int \dd{\tau}\dd{\Sigma} \left(\frac{1}{2}\hat{\lambda}^2 + (m^2-r)\hat{\lambda}\right)\biggr],
 \end{aligned}
\end{equation}
and $V_{\Sigma}$ is the volume of $\Sigma$. 
In the expression above we have introduced the propagator $\Delta^i$ which fulfills
\begin{align}
    (-D^i_\mu D^i_\mu + m^2)\Delta^i(x) = \frac{1}{\sqrt{\det(g)}} \delta(x).
\end{align} 
We will proceed in the same way as in~\cite{Alvarez-Gaume:2019biu} and compute the saddle-point equations where we set the fluctuations $\hat{\lambda}$ to zero. Thus we get 
\begin{align}\label{eq:saddle_point_equations}
	 \eval{\partial_{m^2}S_Q }_{\hat{\lambda} = 0} &= \sum_{i=1}^N \left[V_{\Sigma}\beta \frac{A_i^2}{2}\left(1-\frac{2(m^2-r)}{uA_i^2}\right) + \partial_{m^2}\Tr\log(-D_\mu D^\mu + m^2 )\right] = 0, \\ 
  \eval{\partial_{\theta_j}S_Q}_{\hat{\lambda} = 0} &= \theta_j \frac{V_{\Sigma}A_j^2}{\beta} + \partial_{\theta_j}\Tr\log(-D_\mu D^\mu + m^2 ) - i Q_j = 0, \\ 
	\eval{\partial_{A_j}S_Q }_{\hat{\lambda} = 0} &= V_{\Sigma}\beta A_j\left(\frac{\theta_j^2}{\beta^2} + m^2\right) = 0.
\end{align} 
We want to consider the limit $\beta \to \infty$.
The functional integral can be expressed in terms of zeta functions:
\begin{align}\label{eq:zeta_function_reg_from_saddle_I}
    \partial_{m^2}\Tr\log(-D_\mu D^\mu + m^2 ) = \zeta(1|\theta,\Sigma,m),
\end{align} 
where $\zeta(s|\theta,\Sigma,m)$ is 
\begin{align}
  \zeta(s|\theta,\Sigma,m) = \sum_{n \in \mathbb{Z}}\sum_p \left(\left(\frac{2\pi n}{\beta}+ \frac{\theta}{\beta}\right)^2 + E(p)^2 + m^2\right)^{-s}
\end{align} 
and $E(p)^2$ are the eigenvalues of the Laplacian on the manifold $\Sigma$.
In the \(\beta \to \infty\) limit,
\begin{equation}
    \zeta(1|\theta,\Sigma,m) \underset{\beta \rightarrow \infty}{\longrightarrow } \frac{\beta}{2}\zeta(\sfrac{1}{2}|\Sigma,m).
\end{equation}
From the second and the third saddle-point equations we get
\begin{align}
	\theta_j &=\theta = im \beta,  & A^2_j = \frac{Q_j}{mV_{\Sigma}}.
\end{align}
In terms of the charge $Q = \sum_{i=1}^N Q_i$ the first line becomes
\begin{equation}\label{eq:saddle_point_beta_infty_charge_N}
	\frac{Q}{N} = m \left(\frac{2V_{\Sigma}}{u}(m^2-r)-\zeta(\sfrac{1}{2}|\Sigma,m)\right).
\end{equation} 
To compute the free energy at the saddle we use
\begin{equation}
    \Tr\log(-D_\mu D^\mu + m^2) = - \eval{ \dv{s} \zeta(s|\theta,\Sigma,m) }_{s=0} \underset{\beta \rightarrow \infty}{\longrightarrow} \beta \zeta(-\sfrac{1}{2}|\Sigma,m)
\end{equation}
and find
\begin{equation}
	\label{eq:free_energy_saddle_point}
   \begin{aligned}
     F(Q) &= -\frac{1}{\beta}\sum_{i=1}^N \left[i\theta_i Q_i +\beta \frac{V_{\Sigma}(m^2-r)^2}{2u}- \beta \zeta(-\sfrac{1}{2}|\Sigma,m)\right] \\ 
     &= mQ -\frac{NV_{\Sigma} (m^2-r)^2}{2u}+ N\zeta(-\sfrac{1}{2}|\Sigma,m),
   \end{aligned}
 \end{equation}
where $m$ is to be taken at the saddle point.

\bigskip
The large-N limit is the thermodynamical limit and the saddle-point equations realize the standard thermodynamical relations. We can therefore introduce the grand potential density
\begin{equation}\label{eq:grandPotential}
	\omega(m) = \lim_{\beta\to\infty} -\frac{1}{\beta V_{\Sigma}}\log(Z_{gc}(m)) = 2N\left(-\frac{1}{2V_{\Sigma}}\zeta(-\sfrac{1}{2} |\Sigma, m ) +\frac{(m^2-r)^2}{4u}\right),
\end{equation}
so that the free energy density $f=F/V_{\Sigma}$ is the Legendre transform of $\omega$,
\begin{equation}\label{eq:freeEnergy}
	\begin{cases}
		f = m \rho - \omega(m),\\
		\rho = \dv{\omega}{m}.
	\end{cases}
\end{equation}

\subsection{The effective actions}
\label{sec:EFT}

Having computed the one-loop determinant at fixed charge allows us to write an effective potential thanks to the extra structure that fixing the charge has introduced.

We can construct the 1-loop effective action using thermodynamical reasoning~\cite{Greiter:1989qb}. 
If a physical system is described by a grand potential $\omega=\omega(m)$, one can write an effective microscopic description in terms of a field $\chi$ using the Lagrangian 
\begin{equation}\label{eq:nlsm-chi}
	\mathcal{L}_{\text{NLSM}}=\omega(\abs{\del_\mu\chi \del^\mu\chi}^{1/2}).
\end{equation}
In our case, we use the grand potential computed in Eq.~\eqref{eq:grandPotential} and find a \ac{nlsm} only in terms of the Goldstone field $\chi$.
By construction, the energy of the ground state $\chi=-im \tau$ of this \ac{eft} reproduces the free energy given in Eq.~\eqref{eq:free_energy_saddle_point} computed at large N.\footnote{The imaginary units in $\chi$ and later in $\arg{\varphi_1}$ are due to the Wick rotation to Euclidean space.}
The Lagrangian is to be understood as an expansion around the ground state, so possible fractional powers pose no issue.
The Lagrangian~\eqref{eq:nlsm-chi} has the drawback that, apart from containing fractional powers of the fields, it breaks the \(O(2N)\) symmetry, since it only describes a subsector of the theory.

If we want to describe the full large-N behavior of the system we can write an effective action in terms of N complex fields $\varphi_i$.
Imposing \(O(2N)\)-invariance, it must have the form
\begin{equation}%
  \label{eq:LSM-gen}
  \mathcal{L}_{\text{LSM}} = \sum_{i=1}^N \del_\mu \varphi_i^*\del^\mu\varphi_i - V(\abs{\varphi}).
\end{equation}
This Lagrangian must also reproduce the large-charge behavior, and it turns out that this condition is sufficient to fix the form of \(V(\abs{\varphi})\). 
In other words, the (effective) potential \(V\) is determined by requiring that if we fix one of the U(1) charges, we must recover the initial \ac{nlsm} when eliminating the radial mode.
By construction, the effective potential preserves the global O(2N) symmetry, \emph{i.e.} it depends on $\varphi^\dag \varphi = \abs{\varphi^2}$, so we can choose to fix any of the \(U(1)\) charges, say the one that rotates the field $\varphi_1$.
We make the fixed-charge ansatz 
\begin{align}
	\abs{\varphi_1} &= \Phi, & \arg(\varphi_1) &= -im\tau.
\end{align}
The Lagrangian evaluated on this ansatz is given by
\begin{equation}\label{eq:L-LSM}
  \mathcal{L}_{\text{LSM}} = \Phi^2 m^2 - V(\Phi)
\end{equation}
and the \ac{eom} for the radial mode is
\begin{equation}
  m^2 - \dv{(\Phi^2)} V = 0.
\end{equation}
Its solution is a function \(\Phi = \Phi(m^2)\).
Plugging this back into the Lagrangian~\eqref{eq:L-LSM} we must recover
\begin{equation}
  \eval{\Phi^2 m^2 - V(\Phi)}_{\Phi = \Phi(m^2)} = \omega(m).
\end{equation}
The last two equations define the Legendre transform of the effective potential $V$ seen as a function of $\Phi^2$.
In the following we will need, however, to use a more general definition of the Legendre transform to take into account the fact that some of these functions are only defined for positive values of the parameters.\footnote{The same generalization is used to take into account possible cusps in the effective potential~\cite{duncan2012conceptual}. For a more detailed discussion of this definition, see also~\cite{other-paper}.}
If we  indicate the Legendre transform as
\begin{equation}
  f^*(y) = \sup_{x }( x y - f(x)),
\end{equation}
and introduce the notation
\begin{align}
  x&= \varphi_i^*\varphi_i, & \Upsilon(x) &= V(\sqrt{x}), & y&= m^2 , & \varpi(y) &= \omega(\sqrt{y}) ,
\end{align}
the two effective actions are related by
\begin{equation}
  V(\abs{\varphi}) = \Upsilon(\abs{\varphi}^2) \to \Upsilon^*(m^2) = \varpi(m^2) = \omega(m).
\end{equation}

The Legendre transform preserves the property of homogeneity in the variables, so that for any constant \(a\),
\begin{align}
	f(x) &= a g(x) \implies  f^*(y) = a g^*(y/a),\\
	f(x) &=g(ax) \implies  f^*(y) = g^*(y/a).
\end{align}
In particular, the free energy density as a function of the charge density has the same functional form as the free energy as a function of the charge.

\section{The $\varphi^4$ model on flat space}
\label{sec:phi4}

\subsection{The effective potential on flat space}
\label{sec:torusEFT}

\begin{figure}
  \centering
  \includestandalone[mode=buildnew]{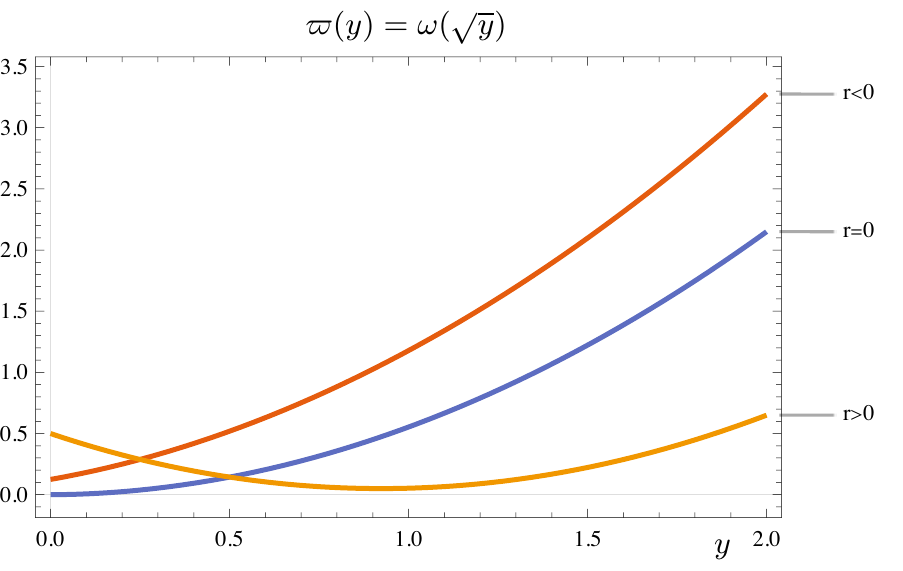}
  \caption{The function \(\varpi(y) = \omega(\sqrt{y})\) for different values of \(r\). Observe that for \(r<0\) the derivative \(\varpi'(y)\) is never smaller than the value reached of \(y = 0\) (\emph{viz.} \(\varpi'(0) = - N r /u \)).}
  \label{fig:Omega}
\end{figure}

We have collected all the ingredients to write down the effective potential at leading order in N.
In the case of the torus, \(\Sigma = T^2\), the grand potential is given by Eq.~\eqref{eq:grandPotential} with the zeta function~\cite{Alvarez-Gaume:2019biu}
\begin{equation}
	\zeta(-1/2|T^2,m) = -\frac{V_{\Sigma}}{6\pi} m^3,
\end{equation}
resulting in
\begin{equation}
	\omega(m) = (2N) \left[\frac{m^3}{12\pi} + \frac{(m^2-r)^2}{4u} \right].
\end{equation}
In principle, this function contains the full information about the model. To make the information more transparent, we follow the construction described in Section~\ref{sec:EFT} and define
\begin{align}
	\varpi(y) &= (2N)\left[\frac{y^{3/2}}{12\pi} + \frac{(y-r)^2}{4u} \right],\\
	\Upsilon(x)& =\varpi^*(x) = \sup_{y\in \setR^+} (xy - \varpi(y)) , 
\end{align}
so that the effective potential is given by
\begin{equation}
	V(\abs{\varphi}) = \Upsilon(\abs{\varphi}^2).
\end{equation}
The function $\varpi(y)$ is convex, so its Legendre transform is well defined.
To find the supremum, we derive w.r.t $y$ at fixed $x$,
\begin{equation}\label{eq:sup}
	x - \varpi'(y) = x - N\left[\frac{y^{1/2}}{4\pi} + \frac{y-r}{u} \right] =0.
\end{equation}
The variable $x$ (which we will identify with $\abs{\varphi}^2$) is defined for $x \in \setR^+$.
There are two possibilities (see Figure~\ref{fig:Omega}):
\begin{itemize}
	\item if $r\geq 0$, the above equation always admits one solution, namely
	\begin{equation}
		y(x) = \pqty{\frac{u}{8\pi}}^2\pqty{1+\sqrt{1+\eta}}^2,
	\end{equation}
	where
	\begin{equation}
		\eta = 64\pi^2\pqty{\frac{x}{Nu} + \frac{r}{u^2}}.
	\end{equation}
	Then
	\begin{equation}
		\Upsilon(x) = \eval{xy - \varpi(y)}_{y=y(x)}
	\end{equation}
and finally 
\begin{equation}\label{eq:effPotTorus}
	V(\abs{\varphi}) = \frac{Nu^3}{3\cdot 2^{10} \pi^4} \left[ 1+\frac{3}{2}\eta +\frac{3}{8}\eta^2 - (1+\eta)^{3/2}\right] - \frac{Nr^2}{2u} =:\hat V(\abs{\varphi}).
\end{equation}
\item If $r<0$, Eq.~\eqref{eq:sup} admits a solution only for $x>-\frac{Nr}{u}>0$.
  For smaller values of $x$, the supremum is obtained for the value of $y$ which minimizes $\varpi(y)$, namely $y=0$:
\begin{equation}
	\Upsilon(x) = \begin{cases}
		\eval{xy-\varpi(y)}_{y=y(x)} & \text{for $x>-\frac{rN}{u}$}\\
		\eval{xy-\varpi(y)}_{y=0}  & \text{for $0<x<-\frac{rN}{u}$.}
	\end{cases}
\end{equation}
For small values of \(x\) the potential is constant, while for big values we recover the same form as above:
\begin{equation}
	V(\abs{\varphi}) = \begin{cases}
		-\frac{Nr^2}{2u} & \text{for $0<\abs{\varphi}<\sqrt{-\frac{rN}{u}}$}\\
		\hat V(\abs{\varphi}) & \text{for $\abs{\varphi} > \sqrt{-\frac{rN}{u}}$.}
	\end{cases}
\end{equation}
\end{itemize}

The regions $r<0$ and $r>0$ are precisely those that correspond to the broken und unbroken phase of the tree-level potential
\begin{equation}\label{eq:tree-level-V}
	V_0 (\abs{\phi}) = r \abs{\phi}^2 +\frac{u}{2N}\abs{\phi}^4.
\end{equation}
These phases are separated by the critical point at $r=0$. We will now study these three cases separately to see how the quantum effects (which have been resummed at leading order in $N$) modify this picture.

\begin{figure}
  \centering
  \includestandalone[mode=buildnew]{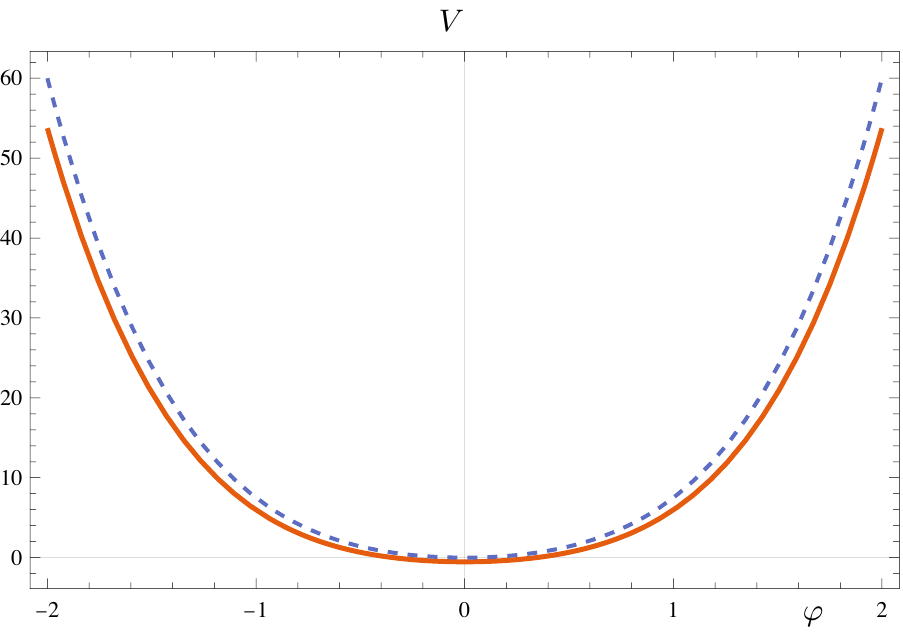}
  \caption{Tree level (dashed line) and effective potential in the unbroken phase.}
  \label{fig:unbroken-potential}
\end{figure}

\paragraph{The unbroken phase, $r>0$.} In the unbroken phase, the behavior of $V$ is qualitatively the same as in the naive tree-level approximation $V_0$ (see Figure~\ref{fig:unbroken-potential}).
The potential is modified to $V(\abs{\varphi}) = \hat V(\abs{\varphi})$ and is given for small values of the field (up to an irrelevant constant) by
\begin{equation}
  \label{eq:V-unbroken-expanded}
  V(\abs{\varphi}) = c + \left[ r + \frac{u^2}{32\pi^2}\left(1 - \sqrt{1+\frac{64\pi^2r}{u^2}}\right)\right]\abs{\varphi}^2 + \frac{u}{2N}\left[1-\frac{1}{\sqrt{1+\frac{64\pi^2r}{u^2}}} \right] \abs{\varphi}^4+\dots
\end{equation}
We can interpret the two coefficients as the values of the renormalized couplings,
\begin{align}
  r & \to r \pqty{1 + \frac{u^2}{32\pi^2 r}\pqty{1 - \sqrt{1+\frac{64\pi^2r}{u^2}}}}, \\
  u & \to u \pqty{1-\frac{1}{\sqrt{1+\frac{64\pi^2r}{u^2}}} }.  
\end{align}
More in general we can describe the \ac{rg} flow of the model in terms of Callan--Symanzik equations satisfied by the effective potential.

\begin{figure}
  \centering
  \includestandalone[mode=buildnew]{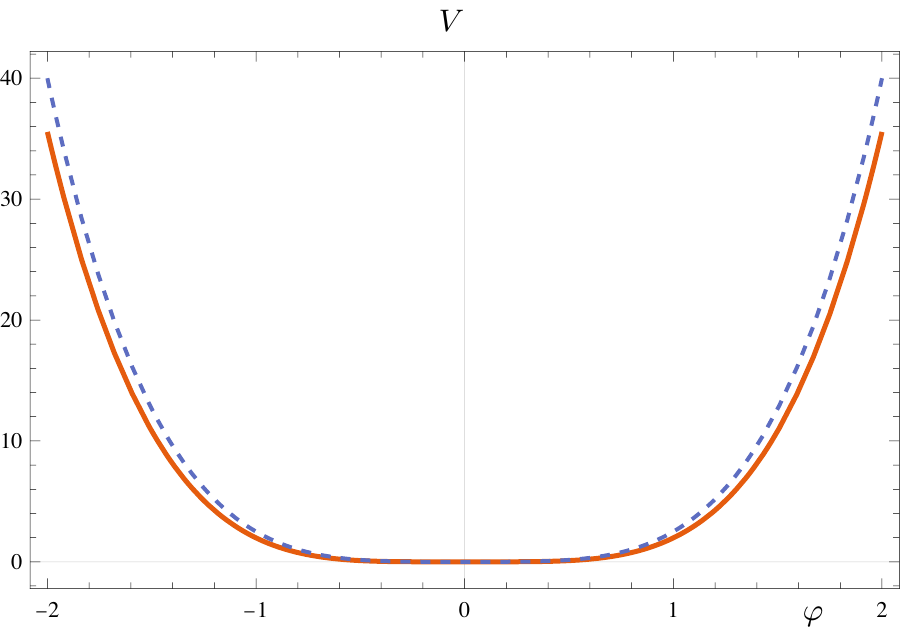}
  \caption{Tree level (dashed line) and effective potential in the critical phase. The potential interpolates between the conformal regime (IR) for large values of \(u\) where \(V \underset{u \to \infty}{\sim} \abs{\varphi}^6\), and the free UV regime for small values of \(u\) where \(V \underset{u \to 0}{\sim} \abs{\varphi}^4\).}
  \label{fig:critical-potential}
\end{figure}

\paragraph{The critical point $r=0$.}
If we fine tune $r=0$, we are at the critical point (see Figure~\ref{fig:critical-potential}).
The effective potential is given by
	\begin{equation}\label{eq:effPotTorusCrit}
	V(\abs{\varphi}) = \frac{Nu^2}{3\cdot 2^{10} \pi^4} \left[ 1 + \frac{3}{2} \times 64\pi^2\frac{\abs{\varphi}^2}{Nu} +\frac{3}{8}\pqty{64\pi^2\frac{\abs{\varphi}^2}{Nu}}^2 - \left( 1 + 64\pi^2 \frac{\abs{\varphi}^2}{Nu}\right)^{3/2}\right],
\end{equation}
agreeing with the result of~\cite{Appelquist:1982vd} based on the resummation of an infinite number of Feynman diagrams, each corresponding to the leading-$N$ contribution to the coefficient of $\abs{\varphi}^n$ in the expansion of the full potential.
We can think of this potential as describing the flow that goes from the Gaussian \ac{uv} fixed point ($u \to 0$) to the strongly coupled \ac{wf} fixed point for $u\to \infty$.
For $u\to 0$ we can expand $V$ in inverse powers of the field,
\begin{equation}
	V(\abs{\varphi}) = \frac{u}{2N}\abs{\varphi}^4\left[1 - \frac{3\pi\sqrt{uN}}{\abs{\varphi}} + \frac{uN}{16\pi^2\abs{\varphi}^2} + \dots\right],
\end{equation}
where each term is again associated to a given Feynman diagram~\cite{Appelquist:1982vd}.

For $u \to \infty$, we have an expansion in positive powers of $\abs{\varphi}$,
\begin{equation}
	V(\abs{\varphi}) = \frac{16\pi^2}{3N^2}\abs{\varphi}^6\left[1 - \frac{24\pi^2\abs{\varphi}^2}{u N} + \frac{4}{3}  \pqty{\frac{24\pi^2\abs{\varphi}^2}{u N}}^2 + \dots\right].
\end{equation}
That the leading term of the potential goes like $\abs{\varphi}^6$ for $u \to \infty$ was to be expected since we are describing a \ac{cft} in terms of an \ac{ir} field of dimension $1/2$.
Since in a \ac{cft}, there are no dimensionful couplings (indeed, $r=0$ and $u\to\infty$), the only way to construct a dimension-three operator is $\abs{\varphi}^6$.
This is precisely the effective potential obtained in the large-charge approximation~\cite{Hellerman:2015nra} and describes the large-charge sector also for finite values of $N$.
We will study it in more detail in the following using the state-operator correspondence and computing also the leading non-perturbative corrections in the large-$N$ approximation.
\paragraph{The broken phase, $r<0$.} The form of the tree-level potential~\eqref{eq:tree-level-V} suggests that for $r<0$, we are in a broken phase realized by the minima of the Mexican hat potential,
\begin{equation}
	\abs{\varphi}^2= \varphi_{\text{min}}^2 = \frac{\abs{r}N}{u}.
\end{equation}
We know however that this picture is changed by quantum corrections: the tree-level potential has a flex, but it is well-known that the effective potential is always a convex function, even in finite volume~\cite{Iliopoulos:1974ur,Israel+2015}. This is precisely what we have found in our calculation above based on the Legendre transform preserving convexity.

\begin{figure}
  \centering
  \includestandalone[mode=buildnew]{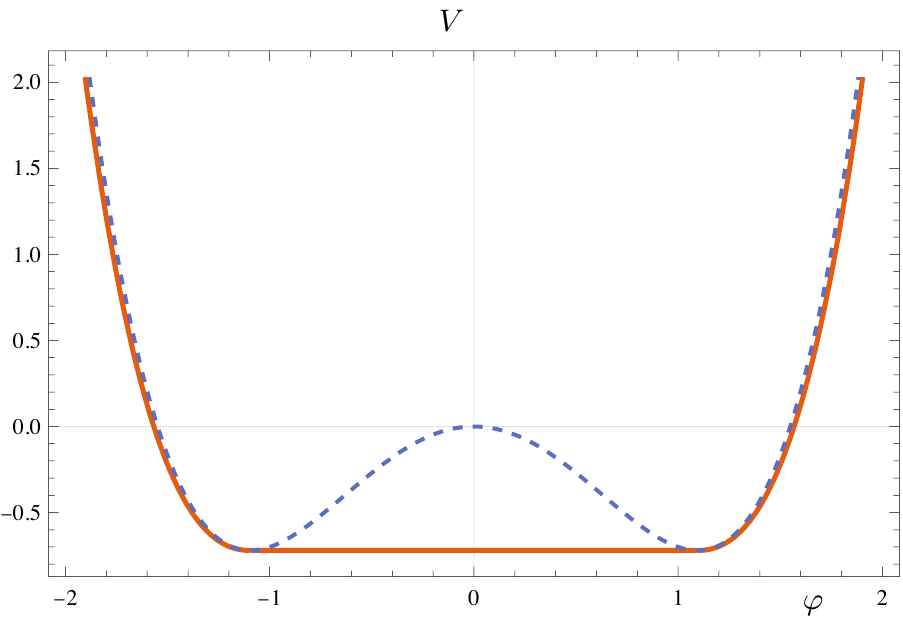}
  \caption{Tree level (dashed) and effective potential in the broken phase. The effective potential is constant in the region between the two minima of the tree-level potential as demanded by convexity.}
  \label{fig:broken-potential}
\end{figure}

Qualitatively, the effective potential $V(\abs{\varphi})$ is close to $V_0(\abs{\varphi})$ in the region of large $\abs{\varphi}$, but constant in the region between the original minima, $\abs{\varphi}^2<\frac{\abs{r}N}{u}$, see Figure~\ref{fig:broken-potential}.
At the minimum, $V(\abs{\varphi})$ has the expansion
\begin{equation}
	V(\abs{\varphi}) = - \frac{r^2N}{2u}+\frac{128\pi^2}{3\sqrt{N}}\left(\frac{\abs{r}}{u}\right)^{3/2}\left(\abs{\varphi}-\sqrt{\frac{\abs{r}N}{u}}\right)^3+\dots
\end{equation}
and has a cusp in the second derivative.

It is natural to wonder what is the actual vacuum of the theory, given that the effective potential is constant for $\abs{\varphi}^2\leq \frac{\abs{r}N}{u}$ and there seems to be no energy argument to prefer any choice in this range.
Moreover, it is very unlikely that this degeneracy can be lifted by higher-order corrections in $N$, as convexity is a nonperturbative property and will in general lead to the same qualitative behavior.

A strong argument for the actual vacuum of the theory to be at the location of the minimum of the tree-level potential (at the extremal value $\abs{\varphi}= \varphi_{\text{min}}$) can be made using cluster decomposition~\cite{duncan2012conceptual}. To illustrate this, consider a simplified version of our model with a single real scalar field, where there is no Goldstone mode and the system has exactly two vacua $\ket{\pm}$, such that%
\footnote{It can be useful to think of the two states \(\ket{\pm}\) as two thermodynamic phases which coexist in the intermediate region (Maxwell's rule).}
\begin{equation}
	\ev{\varphi}{\pm} = \pm \varphi_{\text{min}}.
\end{equation}
Consider now a linear superposition
\begin{equation}
	\ket{\theta} = \cos(\theta) \ket{+} + \sin(\theta) \ket{-}.
\end{equation}
The \ac{vev} of the field on $\ket{\theta}$ covers all the values between $+v$ and $-v$,
\begin{equation}
	\ev{\varphi}{\theta} =  \cos(2\theta) \varphi_{\text{min}}.
\end{equation}
Since $\ket{\theta}$ has zero energy this implies that $V(\abs{\varphi})$ is constant for $-\varphi_{\text{min}}\leq \varphi \leq \varphi_{\text{min}}$~\cite{Weinberg:1987vp}.
The states $\ket{\theta}$ are, however, not in general acceptable vacua.
To see this we can compute the connected two-point function,
\begin{equation}
	\ev{\varphi(x_1)\varphi(x_2)}{\theta}_c = \ev{\varphi(x_1)\varphi(x_2)}{\theta}-\ev{\varphi(x_1)}{\theta}\ev{\varphi(x_2)}{\theta}.
\end{equation}
In the limit $|x_1-x_2| \to \infty$ we find
\begin{equation}
	\ev{\varphi(x_1)\varphi(x_2)}{\theta}_c = 2 \varphi_{\text{min}}^2 \sin(\theta)\cos(\theta) + e^{-|x_1-x_2|}
\end{equation}
which vanishes, as it should, only for $\cos(\theta)=0$ or $\sin(\theta)=0$, \emph{i.e.} for one of the two original minima.

\subsection{The free energy density on flat space}
\label{sec:freeEnergyFlat}

The grand potential \(\omega(m)\) was originally introduced in~\cite{Alvarez-Gaume:2019biu} to compute the free energy density (at fixed \(U(1)\) charge) at the critical point.
Since we have now the expression for \(\omega(m)\) in the general case, we can repeat the computation, which amounts to a Legendre transform of \(\omega(m)\), this time with respect to the variable \(m\):
\begin{equation}
  f(\rho) = \omega^*(\rho) = \sup_{m \in \setR^+} ( m \rho - \omega(m)).
\end{equation}
Note that there is no convexity constraint on \(\omega(m)\) and in fact,
\begin{align}
  \omega''(m_{\text{fl}}) &= 0 , & \text{for\quad \(m_{\text{fl}} = \frac{u}{12 \pi} \pqty{\sqrt{1+ \frac{48 \pi^2 r}{u^2} } - 1} \).}
\end{align}
For \(r \le 0\), \emph{i.e.} in the broken phase and at the critical point, \(\omega\) is convex, while there is a flex for positive values of \(m\) in the massive \(r> 0\) case.

The maximization condition \(\rho = \omega'(m)\) admits a solution for any real positive value of \(\rho\), independently of the sign of \(r\).
What changes is that in the massive case, the region \(m < m_{\text{fl}}\) does not contribute to the Legendre transform.
In other words, if we look at the inverse Legendre transform, there are regions of \(m\) that cannot be reached starting from a fixed-charge description.
This is not surprising: in general we expect the fixed-chemical potential and fixed-charge regimes to be different (see \emph{e.g.} the discussion in the appendix of~\cite{Gaume:2020bmp}).

Explicitly, we can solve the maximization in terms of trigonometric functions:
\begin{equation}
  m(\rho) = \frac{u}{12 \pi} \pqty{ 2 \sqrt{1 + \frac{48 \pi^2 r}{u^2} } \cos(\frac{\theta}{3} ) - 1},
\end{equation}
where
\begin{equation}
  \cos(\theta) = \pqty{1 + \frac{48 \pi^2 r}{u^2}}^{-3/2} \pqty{ \frac{432 \pi^3 \rho}{N u^2} - \pqty{1 + \frac{72\pi^2 r}{u^2} }}.
\end{equation}
Note
that the function is real and positive for any value of \(\rho\) since the function \(\cos(\frac{1}{3}\arccos(\theta) )\) can be continued analytically for \(\theta > 1\).
The resulting free energy density is
\begin{multline}
  f(\rho)  = \frac{N u^3}{2^9 3^4 \pi^4}  \Bigg[ 4 \pqty{ 1 + \frac{144 \pi^2 r}{u^2} } + 3 \pqty{1+ \frac{48 \pi^2 r}{u^2} }^2 \bqty{-1 + 4 \cos(\frac{2 \theta}{3} ) + 2 \cos(\frac{4 \theta}{3} )}\\
  - 8 \pqty{1 + \frac{48 \pi^2 r}{u^2} }^{3/2} \cos(\theta) \Bigg].
\end{multline}

To understand the actual physics it is convenient to expand the free energy density for small values of the charge, using the identities
\begin{align}
  f(0) &= - \omega(m_{\text{min}}), & f'(0) &= m_{\text{min}}, & f''(0) &= \frac{1}{\omega''(m_{\text{min}})} .
\end{align}
In this way we can identify typical behaviors that characterize the possible phases.

\paragraph{Broken phase.}
In the broken phase (\(r< 0\)), we find
\begin{equation}
  f_{\text{br}}(\rho) = - \frac{N r^2}{2u} + \frac{u}{4 N \abs{r}} \rho^2 + \frac{u^3}{48 N^2 \pi r^3} \rho^3 - \frac{u^3}{32 N^3 r^4} \pqty{1 + \frac{u^2}{8 \pi^2 r}} \rho^4 + \order{\rho^5}.
\end{equation}
Apart from the unphysical constant, at leading order the energy depends on the \emph{square of the charge density}, as in the simple mechanics example of the rigid rotor.

\paragraph{Unbroken phase.}
In the unbroken phase (\(r > 0\)) we find
\begin{multline}
  f_{\text{unb}}(\rho) = \frac{N u^3 }{3 \times 2^{10} \pi ^4} \left(1 + \frac{96 \pi ^2 r}{u^2} - \left(1 +\frac{64 \pi ^2 r}{u^2}\right)^{3/2} \right) + \frac{u}{8 \pi} \pqty{\sqrt{1+\frac{64 \pi^2 r}{u^2} } - 1} \rho \\
+\frac{16 \pi^2}{N u} \frac{1}{1 + \frac{64 \pi^2 r}{u^2} - \sqrt{1 + \frac{64 \pi^2 r}{u^2}} }  \rho^2 + \order{\rho^3} ,
\end{multline}
which at lowest order is \emph{linear in the charge density} and consistent with a  particle of mass
\begin{equation}
  m^2= r \pqty{1 + \frac{u^2}{32\pi^2 r}\pqty{1 - \sqrt{1+\frac{64\pi^2r}{u^2}}}} , 
\end{equation}
which is the coupling of \(\varphi^2\) in the effective potential in Eq.~(\ref{eq:V-unbroken-expanded}).

\paragraph{Critical phase.}
For \(r = 0\) we find
\begin{equation}
  f_{\text{crit}}(\rho) = \frac{2 \sqrt{2 \pi } \rho^{3/2}}{3 \sqrt{N}}-\frac{2 \pi ^2 \rho^2}{N u}+\frac{8 \sqrt{2} \pi ^{7/2} \rho^{5/2}}{N^{3/2} u^2}-\frac{256 \pi ^5 \rho^3}{3 \left(N^2
   u^3\right)}+\frac{528 \sqrt{2} \pi ^{13/2} \rho^{7/2}}{N^{5/2} u^4} + \order{ \frac{\rho^4}{u^5}}.
\end{equation}
In particular, at the critical point \(u \to \infty\) the free energy is proportional to \(\rho^{3/2}\), consistently with the predictions of the large-charge expansion in~\cite{Hellerman:2015nra,Loukas:2018zjh}.
In this case the system has an intermediate behavior, midway between the rotor (\(\propto \rho^2\)) and the massive particle (\(\propto \rho\)).
The simplest explanation is that in the critical limit scale invariance fixes the relationship between the energy density (dimension three) and the charge density (dimension two).

\subsection{Nonperturbative finite size corrections on flat space}
\label{sec:NonperturbativeFlat}

Up to this point we have neglected non-perturbative corrections. On the torus, we can however do better as the zeta function is known exactly. The exponential terms in the zeta function can be understood as exponentially suppressed finite-size effects contributing to the effective potential.

The expression that we have used for the zeta function on the torus receives exponentially suppressed corrections in \(m\).
They have been computed \emph{e.g.} in~\cite{Dondi:2021buw} and we can use them to write down the exact expression of the grand potential \(\omega(m)\) at leading order in \(N\).
For a torus of side \(L\), we have
\begin{equation}
\begin{aligned}
  \frac{\omega(m)}{2N} &= - \frac{1}{2 L^2} \zeta(-\sfrac{1}{2}|T^2, m) + \frac{\pqty{m^2 - r}^2}{4u}\\
  & = \frac{m^3}{12 \pi} \pqty{ 1 + \sideset{}{'}\sum_{\mathbf{k}}\frac{e^{-\norm{\mathbf{k}} m L }}{\norm{\mathbf{k}}^2 m^2 L^2 } \pqty{1 + \frac{1}{\norm{\mathbf{k}} m L} }  } + \frac{\pqty{m^2 - r}^2}{4u},
\end{aligned}
\end{equation}
where the sum runs over a square lattice of length \(1\) minus the origin: \(\mathbf{k} = \set{(k_1, k_2)|k_1 = 0,1,\dots; k_2 = 0,1, \dots; (k_1,k_2) \neq (0,0)}\).
The exponential terms have a geometric interpretation and can be understood in terms of massive particles propagating along the cycles of the torus~\cite{Dondi:2021buw}.

One can in principle take into account all the exponential corrections, but for ease of exposition we will keep only the first one:
\begin{equation}
  \frac{\omega(m)}{2N} =  \frac{ m^3}{12 \pi} \pqty{ 1 + 2 \frac{e^{- m L }}{m^2 L^2 } \pqty{1 + \frac{1}{ m L} } + \order{e^{-\sqrt{2}m L}}  } + \frac{\pqty{m^2 - r}^2}{4u}.
\end{equation}
It is convenient to think of this this expression as a trans-series and compute the Legendre transform of \(\varpi(y)\) separately for the exponentially suppressed term.%

The final result is that the leading finite-size correction to the \(\order{1}\) potential that we have computed in Section~\ref{sec:torusEFT} is given by
\begin{equation}
  V(\varphi) = \hat V(\varphi) -\frac{N}{3 \pi L^3} e^{- \frac{u L}{8\pi}  (1 + \sqrt{1+\eta})}\pqty{1+\frac{4u \left(3+\eta +3 \sqrt{1+\eta }\right) L + u^2 \left(2+\eta +2 \sqrt{1+\eta }\right) L
   ^2  }{32 \pi ^2 \left(2+\sqrt{1+\eta }\right)}},
\end{equation}
where again
\begin{equation}
  \eta = 64\pi^2\pqty{\frac{\abs{\varphi}^2}{Nu} + \frac{r}{u^2}}.
\end{equation}
This correction preserves the convexity of the potential.%

\section{The effective potential on the sphere}
\label{sec:sphereEFT}

We can use the same approach as before for constructing an effective action in curved space. We specialize in this section to the sphere as this allows us to invoke the state-operator correspondence at the conformal point and directly read off the scaling dimension of the lowest operator of a given charge $Q$ from the free energy $F(Q)$.

In this case the dimensionful parameter is the curvature of the sphere. 
There is an important technical difference \emph{w.r.t.} flat space since the large-charge \ac{nlsm} is only known perturbatively as an expansion in $R/\rho$. 
Also, the effective potential can only be written perturbatively and it is not clear whether it is possible to resum it in terms of elementary functions.
The \ac{nlsm} Lagrangian at the critical point $r=1/4$, $u\to \infty$ on a sphere of radius $r_0$ is given by
\begin{equation}
  \mathcal{L}_{\text{NLSM}} = \frac{1}{3}\abs{\del_\mu \chi \del^\mu \chi}^{3/2} - \frac{1}{6 r_0^2}\abs{\del_\mu \chi \del^\mu \chi}^{1/2} + \frac{1}{60 r_0^4} \abs{\del_\mu \chi \del^\mu \chi}^{-1/2} + \frac{17}{15360 r_0^6}\abs{\del_\mu \chi \del^\mu \chi}^{-3/2}+\dots
\end{equation}
We can think of it as a perturbative expansion in $1/r_0^2$, or equivalently in terms of the curvature $R=2/r_0^2$. Similarly, we can think of the effective potential as an expansion 
\begin{equation}
  V(\varphi) = (\varphi_i^*\varphi_i)^3 \sum_{l=0} \gamma_l y^l  
\end{equation}
in terms of the
 dimensionless parameter 
\begin{equation}
  y = \frac{R N^2}{96 \pi^2 (\varphi_i^*\varphi_i)^2}.   
\end{equation}
By performing the Legendre transform term by term,
\begin{equation}
	V(\varphi) = \mathcal{L}_{\text{NLSM}}^*(\varphi_i^*\varphi_i),
\end{equation}
we find
\begin{equation}
\begin{aligned}
	V(\varphi) &= \frac{16\pi^2(\varphi_i^*\varphi_i)^3}{3N^2} \bqty{1 + 3 y - \frac{3}{20} y^2 + \frac{53}{2\,560}y^3 + \dots}\\
	&=\frac{16\pi^2}{3N^2}(\varphi_i^*\varphi_i)^3 + \frac{R}{6}\varphi_i^*\varphi_i - \frac{N^2 R^2}{11\,520\pi^2}(\varphi_i^*\varphi_i)^{-1} + \frac{53N^4 R^3}{424\,673\,280\pi^4}(\varphi_i^*\varphi_i)^{-3} + \dots
\end{aligned}
	\end{equation}
The leading term is the same as in flat space since it is curvature-independent. The linear coupling to the curvature may be surprising as it is not the usual conformal coupling of $R/8$. This is, however, the value found when expanding around the Gaussian fixed point. Here, we expand around a strongly coupled fixed point. Since the value of this coupling is not protected by any symmetry, there is no reason to suppose that the value does not change when expanding around different points. 

The free energy, which at the critical point directly corresponds to the conformal dimension of the lowest operator of a given charge, was computed using the large-$N$ techniques in~\cite{Alvarez-Gaume:2019biu}.

\section{Conclusions}
\label{sec:Conclusions}

In this work we have computed the large-N effective potential of the \(O(2N)\) vector model in three dimensions in the presence of quadratic and quartic interactions.
To do so, we have taken advantage of the double-scaling limit $N\to \infty$, $Q\to \infty$, $Q/N$ fixed, in which we can compute the grand potential, which is related via a Legendre transform to the effective potential to leading order in N.
This potential is independent of the charge fixing and describes the full theory and not just a subsector.
It is equivalent to the resummation of an infinite number of Feynman diagrams.
Depending on the sign of the coefficient of the quadratic term in the \ac{uv} action, the model can be in different phases --unbroken, critical point, broken --  but the effective potential always remains convex thanks to the properties of the Legendre transform.\\
Our results can be naturally extended to a general number of dimensions.
This is done in the companion paper~\cite{other-paper}, where the convexity $\omega$ serves as a criterion to identify non-unitary theories.

\bigskip
One limitation of our analysis is its reliance on the double scaling limit \(N \to \infty\), \(Q \to \infty\) with \(Q/N\) fixed.
It would be very interesting to generalize it to finite values of N, possibly starting from a \ac{nlsm} whose form is dictated by the symmetries of the problem and justifying a semiclassical analysis using the large-charge limit.
In this sense, the grand potential that we have derived here could be the starting point for the exploration of the phase space of more general systems.

\subsection*{Acknowledgements}

 We would like to thank Luis Álvarez-Gaumé, Simeon Hellerman, Igor Pesando and Uwe--Jens Wiese for enlightening discussions and Rafael Moser for collaboration on a related project.
 The work of S.R. is supported by the Swiss National Science Foundation under grant number 200021 192137.
 D.O. acknowledges partial support by the \textsc{nccr 51nf40--141869} ``The Mathematics of Physics'' (Swiss\textsc{map}).

\appendix

\setstretch{1}

\printbibliography{}

\end{document}